# Non-uniform B-spline optimization method for generating swept surfaces

1st Xiaoyan Kui
School of Computer Science and Engineering
Central South University
Changsha, Hunan 410083, China
xykui@csu.edu.cn

2nd Min Yang
School of Computer Science and Engineering
Central South University
Changsha, Hunan 410083, China
yangmin@csu.edu.cn

3rd Songpeng Yao
School of Computer Science and Engineering
Central South University
Changsha, Hunan 410083, China
234712233@csu.edu.cn

4th Hao Wang
School of Computer Science and Engineering
Central South University
Changsha, Hunan 410083, China
haowang123@csu.edu.cn

5th Enya Shen
School of Software
Tsinghua University
Beijing 100084, China
shenenya@tsinghua.edu.cn

6th Qinsong Li
Big Data Institute
Central South University
Changsha, Hunan 410083, China
qinsli.cg@csu.edu.cn

7th Beiji Zou*
School of Computer Science and Engineering
Central South University
Changsha, Hunan 410083, China
bjzou@csu.edu.cn

***Abstract*—Swept surface construction is widely used in computer-aided design. We propose a novel optimization method using non-uniform B-splines to improve the approximate accuracy of swept surfaces. First, discrete points on the swept shape are computed, and geometric properties such as surface area, discrete curvature, first-order derivatives, and their rotation angles are used to derive a distribution function representing surface irregularity, with weights adjusted from samples. Then, feature points are selected based on the distribution function to determine control points for the approximate non-uniform B-spline surface via inverse calculation, producing an optimized approximation. Finally, the number of feature points is adjusted based on the estimated approximation error. Experiments on 969 randomly generated sweep samples and 1 pipe example show that the proposed algorithm achieves similar accuracy with fewer control points, reducing them by about 15.85% at a specified accuracy of 0.01. Moreover, with ample sampling points, it reduces the average error by approximately 51.35% when the feature point multiple is 10 times the path control points, outperforming comparable methods.**

## I. INTRODUCTION

Sweeping is the process of moving a cross-section curve along a path. The surface formed by the collection of all positions of the cross-section curve during sweeping is called the swept surface. Due to its simplicity and intuitive nature, which facilitates designers in converting three-dimensional surface design problems into curve design problems, sweep technology has become one of the most commonly used solid modeling methods[1], widely applied in fields such as shipbuilding, automotive design, and CNC machining of precision parts.

With the widespread application of non-uniform rational B-splines (NURBS) in computer-aided design(CAD), when both the cross-section curve and path are represented by NURBS, swept surfaces often cannot be accurately modeled using NURBS[2]. Consequently, obtaining efficient NURBS approximations for swept surfaces has become a key research focus. For general swept surfaces, Coquillart [3] used an offset method to translate path control points to each cross-section control point, obtaining a NURBS approximation. However, for complex paths, the surface control points computed from the path's original control points struggle to achieve global fine control, often resulting in low accuracy of the swept surface. Piegl et al.[2] achieved sweeping surface approximation by placing section curves at each path control point and back-calculating the control points. They also provided a method for increasing control points based on path knot interval subdivision. By continuously subdividing path knot intervals, sweep surface approximations of arbitrary accuracy can be obtained. However, this method is susceptible to the influence of the underlying parameterization technique and considers only path knot information during control point augmentation.

The shape characteristics of the surface can guide the selection of sampling points. Pagani et al.[4] proposed a sampling method based on surface area and curvature, applying it to the approximation of swept surfaces. Inspired by this method, we focus on the high-precision approximation of swept surfaces. Based on the geometric features of the swept surface, we select feature points and propose a non-uniform B-spline optimization method for generating swept surfaces to achieve a higher-precision approximation of swept surfaces.

## II. Fundamentals Of Swept Surface Approximation

The swept surface generated by the cross-section curve $C(u)$ along the path $T(v)$ is defined as follows:

$$S(u,v)=T(v)+\boldsymbol{M}(v)C(u), \quad (1)$$

Where $\boldsymbol{M}(v)$ is a $3\times3$ transformation matrix is used to control the transformation rules during the sweeping of the cross-section curve along the path.

When the cross-section curve $C(u)$ and the path $T(v)$ are represented by NURBS, since the transformation matrix $\boldsymbol{M}(v)$ usually cannot be accurately represented by NURBS[2], it is necessary to calculate the NURBS approximation surface $\hat{S}(u,v)$ of the swept surface $S(u,v)$. The basic process for approximating a swept surface using NURBS surfaces is as follows: First, select feature points on the path, then place the cross-section curve at each feature point on the path; finally, obtain the control points of the approximate surface by solving the linear equation system, thus obtaining the approximate surface. This process will be described in detail below.

### A. Feature point selection.

The accuracy of the approximate surface increases with the increase in the number of feature points. A simple method is to use the midpoint of the longest knot interval of the path as the newly inserted knot value to subdivide the knot interval[2], as shown in Figure 1. For simplicity, we call this method the mid-knot method.

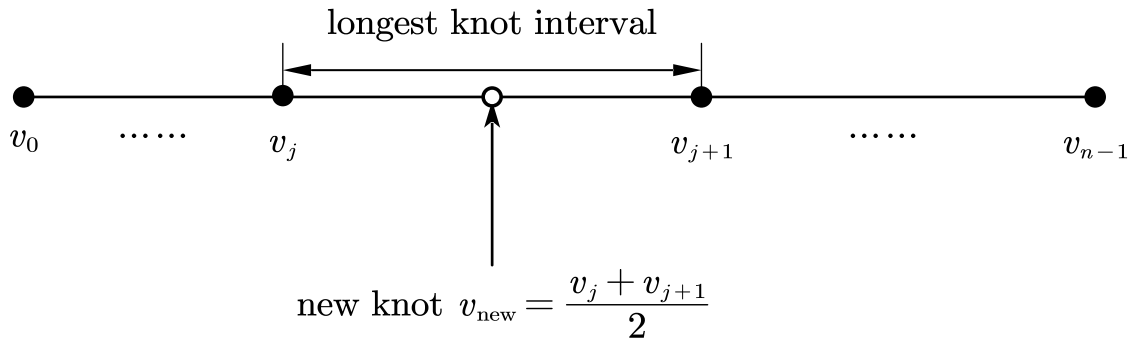

Figure 1. Mid-knot method.

After subdividing the knot intervals, the knot vector $\boldsymbol{V}=\{v_0,\cdots\cdots,v_{\hat{n}+d_v}\}$ in the $v$-direction of the approximate surface is obtained, where $d_v$ is the degree of the path $T(v)$ and $\hat{n}$ is the number of feature points corresponding to the subdivision. Then, the feature point parameter values $\{\bar{v}_i\}_{i=0,\cdots\cdots,\hat{n}-1}$ are calculated using the averaging method:

$$\begin{cases}\bar{v}_0=v_0\\ \bar{v}_i=\dfrac{v_{i+1}+\cdots+v_{i+d_v}}{d_v}, i=1,\cdots,\hat{n}-2.\\ \bar{v}_{\hat{n}-1}=v_{\hat{n}+d_v}\end{cases} \quad (2)$$

When the number of knots is increased sufficiently, the swept surface generated based on the mid-knot method can achieve any specified accuracy. However, this method only focuses on the knot vector information of the path and does not fully consider the relevant information of the swept surface when selecting feature points, resulting in problems such as slow error reduction in practical applications.

### B. Reverse calculation of control points

To simplify the solution process, we only consider the case where the cross-section curve $C(u)$, path $T(v)$, and approximate surface $\hat{S}(u,v)$ are non-uniform B-splines, denoted as:

$$C(u)=\sum_{i=0}^{m-1}N_{i,d_u}(u)P_i^u, \quad (3)$$

$$T(v)=\sum_{j=0}^{n-1}N_{j,d_v}(v)P_j^v, \quad (4)$$

$$\hat{S}(u,v)=\sum_{i=0}^{m-1}\sum_{j=0}^{n-1}N_{i,d_u}(u)N_{j,d_v}(v)P_{ij}, \quad (5)$$

Where $N_{i,d_u}(u)$ and $N_{j,d_v}(v)$ are the B-spline basis functions corresponding to $C(u)$ and $T(v)$, $d_u$ is the degree of the cross-section curve $C(u)$, and $P_i^u$, $P_j^v$, $P_{ij}$ are the control points of $C(u)$, $T(v)$, and $\hat{S}(u,v)$, respectively.

To ensure the accuracy of the isoparametric lines in the $u$-direction of the approximate surface, the control points $\{P_i^u\}$ of the cross-section curve are placed at each feature point $\{\bar{v}_j\}$ of the path according to formula (1). Using the knot vector of the cross-section curve $C(u)$ as the knot vector of the approximate surface $\hat{S}(u,v)$ in the $u$-direction, the initial control points $\{\hat{P}_{ij}^u\}$ are obtained. Substituting it into formula (3) yields the system of linear equations:

$$\begin{pmatrix}\hat{P}_{i,0}^u\\ \vdots\\ \hat{P}_{i,n-1}^u\end{pmatrix}=\begin{pmatrix}N_{0,d_v}(\bar{v}_0) & \cdots & N_{n-1,d_v}(\bar{v}_0)\\ \vdots & \ddots & \vdots\\ N_{0,d_v}(\bar{v}_{n-1}) & \cdots & N_{n-1,d_v}(\bar{v}_{n-1})\end{pmatrix}\begin{pmatrix}P_{i,0}\\ \vdots\\ P_{i,n-1}\end{pmatrix}. \quad (6)$$

The parameter values calculated by equation (2) ensure that the coefficient matrix in equation system (6) is non-singular, and the control points $\{\hat{P}_{ij}\}$ of the approximate surface $\hat{S}(u,v)$ can be obtained by inverse calculation:

$$\begin{pmatrix}\hat{P}_{i,0}\\ \vdots\\ \hat{P}_{i,n-1}\end{pmatrix}=\begin{pmatrix}N_{0,d_v}(\bar{v}_0) & \cdots & N_{n-1,d_v}(\bar{v}_0)\\ \vdots & \ddots & \vdots\\ N_{0,d_v}(\bar{v}_{n-1}) & \cdots & N_{n-1,d_v}(\bar{v}_{n-1})\end{pmatrix}^{-1}\begin{pmatrix}\hat{P}_{i,0}^u\\ \vdots\\ \hat{P}_{i,n-1}^u\end{pmatrix}. \quad (7)$$

### C. Local coordinate frame calculation

Choosing different transformation matrices $\boldsymbol{M}(v)$ can apply different effects to the sweeping process, such as rotating the cross-section curve or keeping the cross-section curve parallel to a specific direction. We only consider the general sweeping process, that is, when the cross-section curve is swept along the path without any additional operations, the transformation matrix $\boldsymbol{M}(v)$ in this case is the rotation minimizing frame (RMF) of the path at $v$. The RMF of the path is defined as[5-6]:

$$\boldsymbol{M}(v)=\left(\boldsymbol{t}(v)^T,\boldsymbol{r}(v)^T,\left(\boldsymbol{t}(v)\times\boldsymbol{r}(v)\right)^T\right). \quad (8)$$

Where $\boldsymbol{t}(v)$ is the tangent vector function of the path $T(v)$, and $\boldsymbol{r}(v)$ satisfies:

$$\begin{cases}\boldsymbol{r}'(v)-\phi(v)\boldsymbol{t}(v)=0\\ \boldsymbol{r}(v)\cdot\boldsymbol{t}(v)=0\end{cases}. \quad (9)$$

Where $\phi(v)$ is any function of $v$, and $\boldsymbol{t}(v)$ and $\boldsymbol{r}(v)$ are both $1\times3$ row vector functions.

The local coordinate frame defined by the RMF does not undergo additional rotation relative to the tangent of the curve and only requires the first derivative of the curve, thus exhibiting favorable properties. Commonly used RMF approximation methods include the projection method[7-8], rotation method[9-11], numerical integration method[12], and double reflection method[5]. We use the double reflection method to calculate the RMF of the path $T(v)$. The double reflection method is based on the frame of the previous point, and obtains the frame of the next point by performing two reflection operations, as shown in Figure 2. For planar curves and spherical curves, the frame calculated by the double reflection method is an accurate RMF. For spatial curves, the double reflection method has a fourth-order global approximation error.

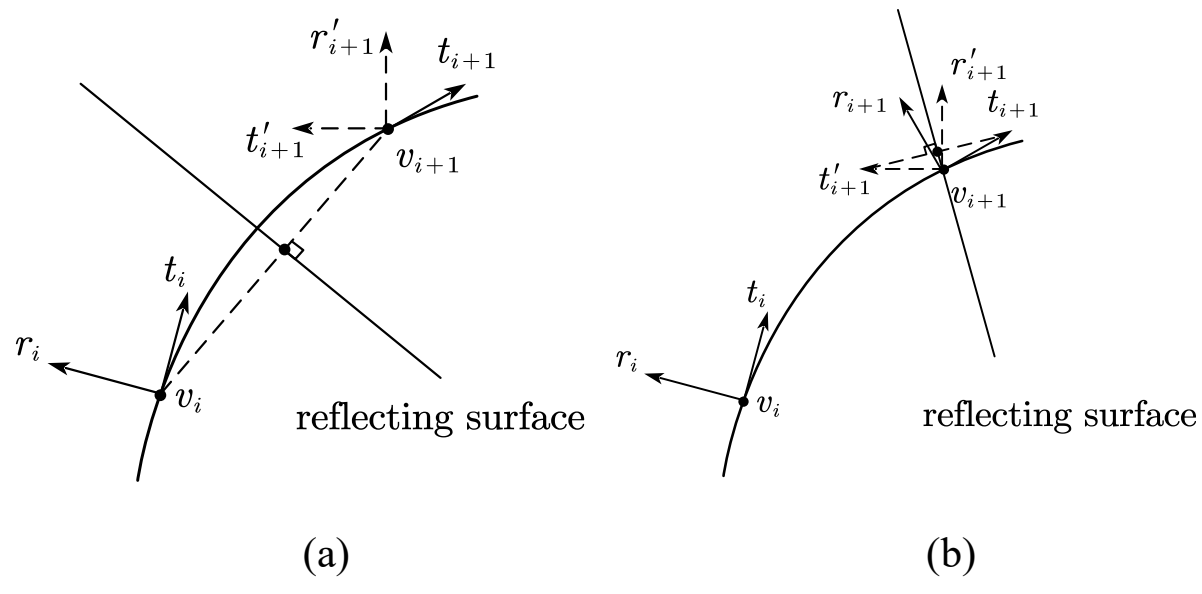

Figure 2. Double reflection method :(a) First reflection; (b) Second reflection.

As shown in equation (1), the calculation error of each point on the swept surface is only affected by the RMF calculation error. By selecting an appropriate frame calculation step size, the calculation error of the swept surface can be ensured to be small enough. It is acceptable to disregard the calculation error of the frame and assume that the position of any point on the swept surface can be precisely calculated.

### D. Error calculation

After obtaining an approximate surface for the swept surface, it is often necessary to estimate the approximation error.

A common method is the Hausdorff distance between surfaces, which requires sampling points in the $u$ and $v$ directions, has a time complexity of $O\left(n_u^2 n_v^2\right)$, and the calculation error is related to the sampling step size. When the approximation accuracy is high, the time required becomes excessive. We introduce an estimation method[13] that converts sweeping surface error into path offset error:

When for $\forall i\in\{0,\cdots,m-1\}$ and $\exists\varepsilon>0$, such that $\left\|T(v)+\boldsymbol{M}(v)P_i^u-\sum_{j=0}^{n-1}P_{ij}N_{j,d_v}(v)\right\|\leqslant\varepsilon$, then:

$$\begin{aligned}S(u,v)-\hat{S}(u,v)&=T(v)+\boldsymbol{M}(v)C(u)-\hat{S}(u,v)\\&=\sum_{i=0}^{m-1}N_{i,d_u}(u)\left[T(v)+\boldsymbol{M}(v)P_i^u-\sum_{j=0}^{n-1}P_{ij}N_{j,d_v}(v)\right]\\&\leqslant\sum_{i=0}^{m-1}N_{i,d_u}(u)\varepsilon=\varepsilon.\end{aligned} \quad (10)$$

This method converts the approximation error of the swept surface into an approximation error based on the offset of control points along the path. It only requires sampling along the path direction and does not require distance calculation between point sets, resulting in a time complexity of $O(n_v)$. We will use this method to estimate the error of the swept surface.

## III. METHODOLOGY

We measure the irregularity of a swept surface by sampling it and approximating its surface area, curvature, first derivative, and first derivative rotation angle. Feature points are selected by calculating the irregularity distribution function of the swept surface in the $v$-direction. The method introduced in Section Ⅱ is then used to approximate the swept surface, and a method to increase the number of feature points is presented to control the accuracy of the approximate surface.

### A. Calculation of swept surface information

The derivative, derivative rotation angle, and curvature can be used to measure the irregularity of a curve[14]. Considering that control points need to be appropriately increased in regions where the surface changes gently, we use surface area, curvature, first derivative, and first derivative rotation angle to measure the irregularity of the swept surface.

Since the RMF obtained by approximate calculation is usually in discrete form, equation (1) is usually not analytical. We calculate the sampling points $\left\{p_{ij}\right\}$ of the swept surface using the isoparametric sampling method, with corresponding parameter values $\{\hat{u}_i\}_{i=0}^{n_u-1}$ and $\{\hat{v}_j\}_{j=0}^{n_v-1}$. Surface information is calculated based on the sampling points:

(1) The surface area in the $v$-direction, $a_j^v$. Estimates were made using the triangular faceted slice approximation:

$$a_j^v = \sum_{i=0}^{n_u-1} a_{ij}. \quad (11)$$

Where $a_{ij}$ is the approximate surface area between four adjacent sampling points $p_{ij}$, $p_{i+1,j}$, $p_{i,j+1}$, $p_{i+1,j+1}$:

$$a_{ij} = S_{\triangle p_{ij} p_{i,j+1} p_{i+1,j}} + S_{\triangle p_{i+1\,j+1} p_{i,j+1} p_{i+1,j}}. \quad (12)$$

(2) The discrete curvature in the $v$-direction, $\kappa_j^v$. The forward differential approximate derivatives are used to perform the calculations[15]:

$$\kappa_j^v = \frac{1}{n_u}\sum_{i=0}^{n_u-1}\kappa_{ij}. \quad (13)$$

Where $\kappa_{ij}$ is the discrete curvature at the sampling point $p_{ij}$:

$$\kappa_{ij} = \sqrt{\left(\kappa_{ij}^u\right)^2 + \left(\kappa_{ij}^v\right)^2}. \quad (14)$$

$$\kappa_{ij}^u = \frac{\left\| \Delta p_{ij}^u \times \Delta^2 p_{ij}^u \right\|}{\left\| \Delta p_{ij}^u \right\|^3}, \kappa_{ij}^v = \frac{\left\| \Delta p_{ij}^v \times \Delta^2 p_{ij}^v \right\|}{\left\| \Delta p_{ij}^v \right\|^3}. \quad (15)$$

Where $\kappa_{ij}^u$, $\Delta p_{ij}^u$, and $\Delta^2 p_{ij}^u$ are the surface discrete curvature, forward first-order difference, and second-order difference of the sampling point $p_{ij}$ in the $u$-direction, respectively, and $\kappa_{ij}^v$, $\Delta p_{ij}^v$, and $\Delta^2 p_{ij}^v$ are the surface discrete curvature, forward first-order difference, and second-order difference of the sampling point $p_{ij}$ in the $v$-direction, respectively.

(3) The first derivative in the $v$-direction, $\boldsymbol{r}_j^v$. The chord length approximation was used:

$$\boldsymbol{r}_j^v = \frac{1}{n_u}\sum_{i=0}^{n_u-1}\frac{p_{i,j+1} - p_{i,j-1}}{\hat{v}_{j+1} - \hat{v}_{j-1}}. \quad (16)$$

(4) The first order derivative rotation angle in the $v$-direction, $b_j^v$. The first-order derivatives at the two sampling points before and after were used for approximation:

$$b_j^v = \frac{1}{n_u}\sum_{i=0}^{n_u-1} angle\left(\frac{p_{i,j+2} - p_{i,j}}{\hat{v}_{j+2} - \hat{v}_j}, \frac{p_{i,j} - p_{i,j-2}}{\hat{v}_j - \hat{v}_{j-2}}\right). \quad (17)$$

Where $angle(\boldsymbol{\alpha}_1, \boldsymbol{\alpha}_2)$ represents the angle between two vectors $\boldsymbol{\alpha}_1$ and $\boldsymbol{\alpha}_2$.

It should be noted that since the $u$-direction isoparametric lines of the surface calculated based on the swept surface approximation method introduced in Section Ⅱ are all cross-section curves, only the $v$-direction information is considered when calculating the first derivative and the first derivative rotation angle. To simplify the calculation, the $u$-direction surface discrete curvature $\kappa_{ij}^u$ in equation (14) is replaced by the curve curvature of the cross-section curve at the corresponding point.

*B. Calculation of the distribution function*

Our method for feature point selection is to select more points where the surface irregularity is high and fewer points where the surface irregularity is low. The degree of surface irregularity is measured by a distribution function $D(v)$, calculated using the following formula:

$$D(\hat{v}_j) = \frac{\omega_a A_j}{A_{n_v-1}} + \frac{\omega_\kappa K_j}{K_{n_v-1}} + \frac{\omega_r R_j}{R_{n_v-1}} + \frac{\omega_b B_j}{B_{n_v-1}}. \quad (18)$$

Where $A_j$, $K_j$, $R_j$, and $B_j$ are the cumulative surface area, cumulative discrete curvature, cumulative first derivative, and cumulative rotation angle of the first derivative in the $v$-direction, respectively, and the calculation formula is:

$$A_j = \sum_{i=0}^{j} a_i^v, K_j = \sum_{i=0}^{j}\kappa_i^v, R_j = \sum_{i=0}^{j}\left\|\boldsymbol{r}_i^v\right\|, B_j = \sum_{i=0}^{j} b_i^v. \quad (19)$$

Parameters $\omega_a$, $\omega_\kappa$, $\omega_r$, and $\omega_b$ represent the degree of attention given to various surface information features when measuring the degree of surface irregularity. A simple method is to assign them all the same weight, i.e., $\frac{1}{4}$. However, the influence of each surface feature on the degree of surface irregularity may differ in different sweep examples. We provide a method for dynamically adjusting the weights:

(1) Calculate the swept approximation surface $\hat{S}(u,v)$ at time $\omega_a = \omega_\kappa = \omega_r = \omega_b = \frac{1}{4}$;

(2) Calculate the error of $\hat{S}(u,v)$ at each sampling point $\{e_{ij}\}$:

$$e_{ij} = \left\| p_{ij} - \hat{S}(\hat{u}_i, \hat{v}_j) \right\|. \quad (20)$$

(3) Adjust the surface information weights $\omega_a$, $\omega_\kappa$, $\omega_r$, $\omega_b$ calculated from sampling point error $\{e_{ij}\}$ and equations (11)-(17):

$$\omega_a = \frac{\frac{1}{n_u}\sum_{j=0}^{n_v-1}\sum_{i=0}^{n_u-1} e_{ij} a_j^v}{\sum_{j=0}^{n_v-1} a_j^v}, \omega_\kappa = \frac{\frac{1}{n_u}\sum_{j=0}^{n_v-1}\sum_{i=0}^{n_u-1} e_{ij}\kappa_j^v}{\sum_{j=0}^{n_v-1}\kappa_j^v},$$

$$\omega_r = \frac{\frac{1}{n_u}\sum_{j=0}^{n_v-1}\sum_{i=0}^{n_u-1} e_{ij}\left\|\boldsymbol{r}_j^v\right\|}{\sum_{j=0}^{n_v-1}\left\|\boldsymbol{r}_j^v\right\|}, \omega_b = \frac{\frac{1}{n_u}\sum_{j=0}^{n_v-1}\sum_{i=0}^{n_u-1} e_{ij} b_j^v}{\sum_{j=0}^{n_v-1} b_j^v}. \quad (21)$$

(4) Normalisation of the weights:

$$\hat{\omega}_a = \frac{\omega_a}{\omega_a+\omega_\kappa+\omega_r+\omega_b}, \hat{\omega}_k = \frac{\omega_\kappa}{\omega_a+\omega_\kappa+\omega_r+\omega_b},$$
$$\hat{\omega}_r = \frac{\omega_r}{\omega_a+\omega_\kappa+\omega_r+\omega_b}, \hat{\omega}_b = \frac{\omega_b}{\omega_a+\omega_\kappa+\omega_r+\omega_b}. \quad (22)$$

By dynamically adjusting the weights for different samples, the calculated distribution function can more accurately reflect the surface characteristics of the samples. The effectiveness of this method was verified in the experiments in Section Ⅳ.D.

*C. Method of selecting feature points*

The selection of feature points follows the principle of uniformly distributing the irregularity of the surface across the selected feature points. Based on the $v$-direction distribution function $D(v)$ calculated in Section Ⅲ.B, when the number of feature points $\hat{n}$ is specified, the sequence of feature points $\{\bar{v}_j\}$ selected from the $v$-direction surface sampling points $\{\hat{v}_j\}$ satisfies:

$$D(\bar{v}_j) - D(\bar{v}_{j-1}) \approx \frac{1}{\hat{n}-1}, \forall j = 1,\cdots,\hat{n}-1. \quad (23)$$

After obtaining the feature points, the average method is used to calculate the $v$-direction knot vector $\{v_j\}$ to ensure that the coefficient matrix in equation (6) is invertible:

$$\begin{cases} v_0 = \cdots = v_{d_v} = 0 \\ v_j = \dfrac{\bar{v}_{j-d_v} + \cdots + \bar{v}_{j-1}}{d_v}, j = d_v+1,\cdots,\hat{n}-1. \\ v_{\hat{n}} = \cdots = v_{\hat{n}+d_v} = 1 \end{cases} \quad (24)$$

Combining the control point inversion method introduced in Section Ⅱ.B, the control points of the swept approximation surface can be obtained, thus yielding the non-uniform B-spline approximation surface. In engineering design, the accuracy $\varepsilon$ of the approximation surface is usually specified, requiring automatic selection of the number of feature points. Based on the method for specifying the number of feature points, we provide a method for automatically selecting the number of feature points to generate an approximation surface with a specified accuracy:

(1) Initialise the number of feature points $\hat{n}$ to equal the number of control points $n$ on the path.

(2) Based on the feature point selection method introduced in equation (23) with the control point inversion method introduced in Section Ⅱ.B, the approximation surface $\hat{S}(u,v)$ is generated;

(3) Select an appropriate step size for path sampling, and calculate an error $e$ between the approximate surface $\hat{S}(u,v)$ and the ideal surface $S(u,v)$ according to equation (10);

(4) If $e > \varepsilon$, then increase the number of feature points $\hat{n} = \hat{n} + n$ and return to step 2 until the specified accuracy is reached or the number of iterations reaches the preset number of iterations.

The number of iterations in this method can be adjusted according to the required surface accuracy. Experiments in Section Ⅳ.C show that when the specified accuracy is 0.1 and 0.01, using an iteration threshold of 20 can satisfy the needs of most examples.

*D. Overall methodology pipeline*

Combining the distribution function calculation method based on surface information, the feature point selection method with specified accuracy in Section Ⅲ, and the control point inversion method in Section Ⅱ, the pipeline of our proposed method for approximating swept surfaces using non-uniform B-splines is summarized as shown in Figure 3.

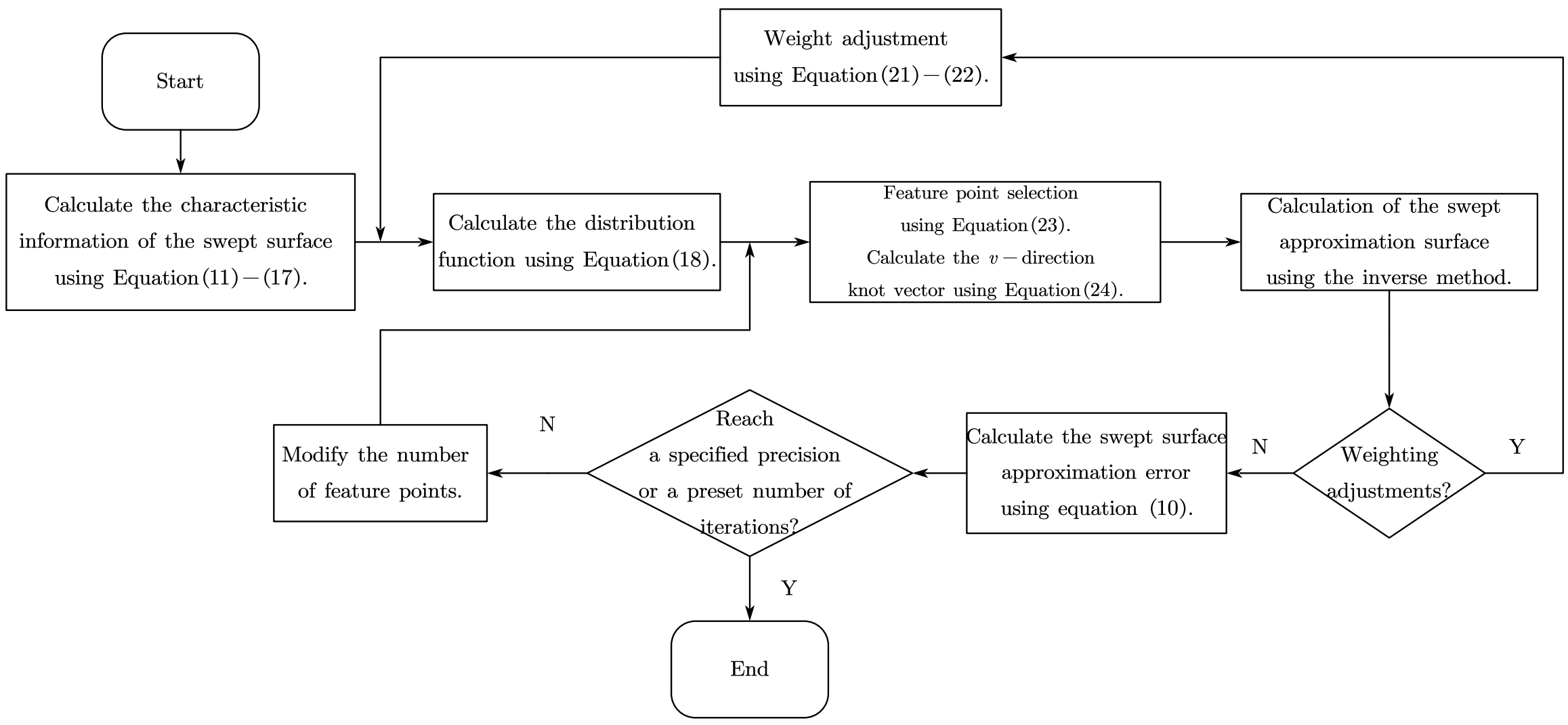

Figure 3. Overall methodology pipeline.

## IV. EXPERIMENT

We use ACIS to design the swept surface. Both the cross-section curve and the path are generated randomly, and ACIS is used to screen the validity of the swept samples. The RMF calculation step size used is 0.01, and the error is calculated using equation (10). The number of path sampling points used is $10^4$, and the number of surface sampling points used to calculate the surface information is $10^3 \times 10^3$. The sampling method is uniform sampling along the parameter direction.

In this section, we first introduce the random generation method of the cross-section curve and path used in this experiment. Based on the randomly generated examples, we verify our method and compare it with similar methods to illustrate the advancement of our method. Additionally, we analyze the effect of the dynamic weight adjustment method separately to illustrate the effectiveness of this module.

### A. Random sample generation method

To verify the applicability of the algorithm in various sweeping examples, the cross-section curves and paths were randomly generated using the method of randomly generating control points and knot vectors. The cross-section curve is in the *xoy* plane with a degree of 3, the number of control points is randomly generated between [4,8], and the coordinate values of the control points obey the distribution $N\left(U\left(-0.3,0.3\right),0.49\right)$. The path has a degree of 3, the number of control points is randomly generated between [4,10], and the coordinates of the *i*th control point obey the distribution $N\left(0.25i,4\right)$. The corresponding knots obey the distribution $U\left(0,1\right)$ except for the first and last knots. Here $N\left(\mu,\sigma^2\right)$ denotes a normal distribution with mean $\mu$ and standard deviation $\sigma$. $U\left(a,b\right)$ denotes a uniform distribution on the interval $(a, b)$.

After randomized samples were subjected to legality checks such as self-intersection and sharp points using ACIS, 969 samples capable of generating swept surfaces were obtained and used for method performance testing. Some of these samples are shown in Figure 4. Due to the randomness of the generation method, these samples can reflect the characteristics of most types of swept surfaces. The proposed method was evaluated based on these random samples to ensure the generalizability and reliability of the experimental results.

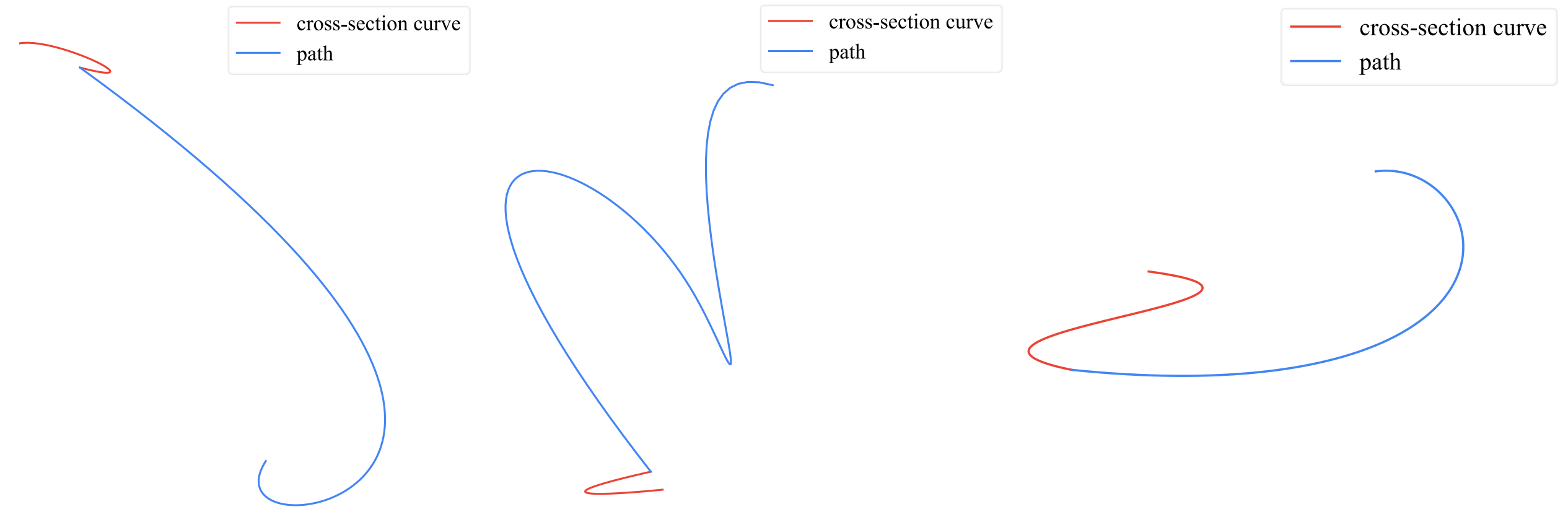

Figure 4. Randomly generated sweep examples (partial)

### B. Analysis of the selected multiplier

Based on 969 randomly generated sweep samples, the error estimation method introduced in equation (10) was used. Our method was compared with the mid-knot method introduced in Section II[2] and a similar method based on curvature and surface area sampling[4].

The average error on 969 samples is shown in Table I, where the selected multiplier refers to the number of selected feature points being a multiple of the number of path control points, and bold indicates the optimal value. When the number of specified feature points is small, the mid-knot method performs better overall. This is because the mid-knot method uses the knot vector of the path to calculate the initial feature point sequence, which can effectively capture the $v$ -direction features when the number of feature points is small. As the number of required feature points increases, our method shows good performance compared with the other two methods. Compared with Pagani et al. method[4], when the multiplier is selected as 7, the average error of our method is reduced by about 13.63%; while when the multiplier is selected as 10, the average error of our method is reduced by about 51.35%. This shows that our method makes more efficient use of surface information.

TABLE I. THE AVERAGE ERROR OF EACH METHOD UNDER DIFFERENT SELECTION MULTIPLES

| Selected multiplier | Mid-knot[2] | Pagani et al[4] | Our method |
|---|---|---|---|
| 1 | **1.074** | 1.492 | 1.644 |
| 3 | **0.377** | 0.527 | 0.774 |
| 5 | **0.218** | 0.273 | 0.321 |
| 7 | 0.168 | 0.154 | **0.133** |
| 10 | 0.110 | 0.074 | **0.036** |

To further illustrate the effectiveness of the proposed method in practical applications, we select a pipe model as an example for analysis, and the model is shown in Figure 5(a). The construction of the curved surface of a pipe is a common application scenario of the sweeping technique, which is usually obtained by sweeping along a path with a closed cross-section curve. The cross-section curve and path corresponding to the example shown in Figure 5(a) are shown in Figure 5(b).

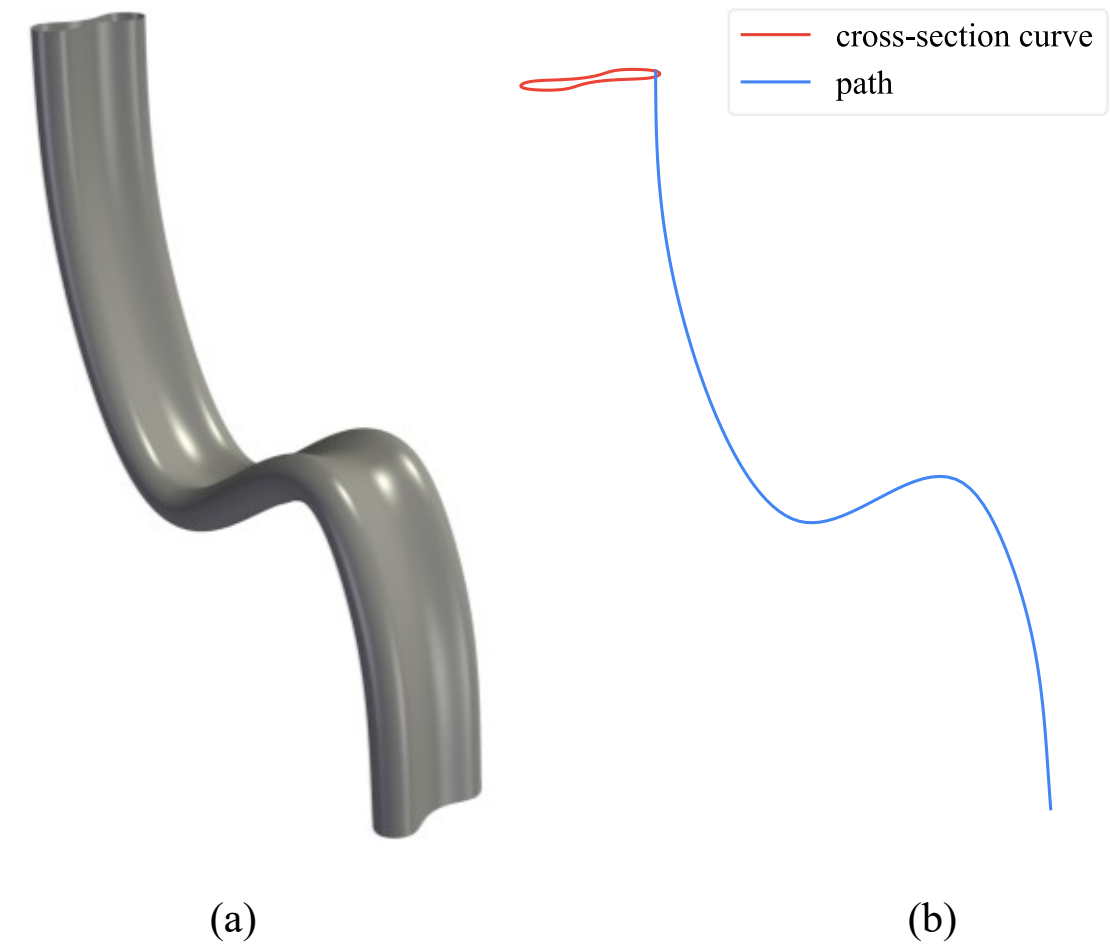

Figure 5. Pipe example: (a) Pipe model example; (b) Cross-section and path corresponding to the pipe example.

Figure 6 shows the error trends of Pagani et al. method[4], and our method on the pipe example under different selection multiples. Similar to the results in Table Ⅰ, the mid-knot method shows better performance when the number of feature points is small, but the rate of error reduction slows down as the number of feature points increases, and convergence is prone to stagnation. This is because the mid-knot method does not utilize the information of the swept surface, making it difficult to capture feature points that affect the surface accuracy during the process of increasing feature points. As the number of feature points increases, our method shows better performance than the other two methods, indicating that our method can more effectively utilize surface information during the selection of feature points.

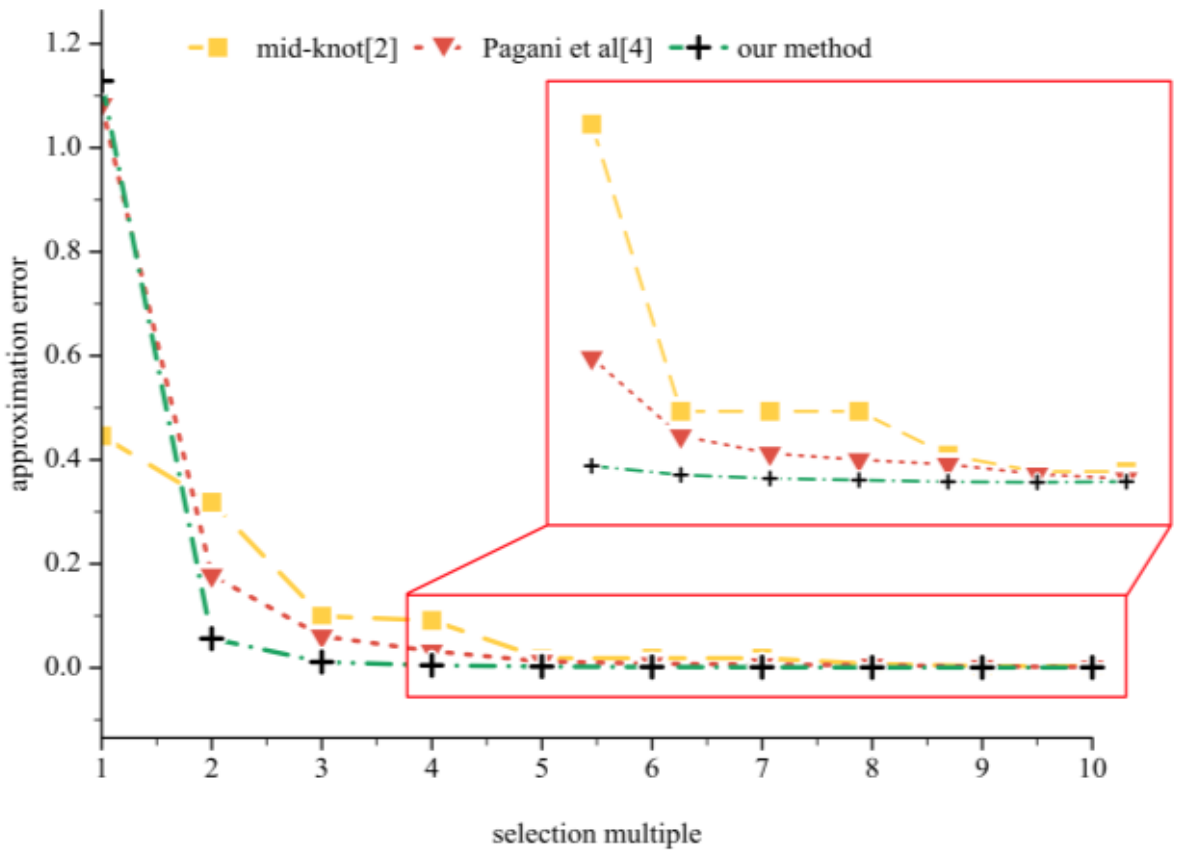

Figure 6. The error of each method on the pipe example under different selection multiples.

Figure 7 shows the error distribution of each method on the example when the selected multiplier is 10, corresponding to 90 feature points. Both the mid-knot method and Pagani et al. method[4] showed large errors at curved surface sections. Figure 8 shows the distribution of cross-section curve placement positions for each method when the selected multiple was 10. When using the same number of feature points, the mid-knot method selected more feature points for cross-section curve placement at the left bend of the path. Pagani et al. method[4] instead selected feature points for placement approximately uniformly along the path. Compared to these two methods, our method selected more feature points for segment placement at the two highly irregular curved segments of the surface, thereby achieving lower error. When using the same number of feature points, our method effectively captures the irregular features on the surface, accurately selects feature points, and consequently generates a high-accuracy approximate surface.

### C. *Analysis of specified accuracy*

To verify the effectiveness of the algorithm under different approximation accuracy requirements, we tested the performance of Pagani et al. method[4], and our method at specified approximation accuracies of 0.1 and 0.01 based on 969 randomly generated sweep samples. The approximation accuracy in this experiment refers to the approximation accuracy of the sweep surface calculated by equation (10). Since neither the mid-knot method nor Pagani et al. method[4] provides a corresponding method for selecting the number of feature points, we adopted the method for selecting the number of feature points proposed in Section Ⅲ.C. The results are shown in Table Ⅱ.

TABLE II. THE AVERAGE MULTIPLES OF FEATURE POINTS REQUIRED BY EACH METHOD UNDER DIFFERENT ACCURACY REQUIREMENTS

| Specified accuracy | Mid-knot[2] | Pagani et al[4] | Our method |
|---|---|---|---|
| 0.1 | 5.239 | 4.736 | **4.055** |
| 0.01 | 7.657 | 7.570 | **6.370** |

As shown in Table Ⅱ, when the specified approximation accuracy is 0.1, the number of feature points required by our method is reduced by approximately 22.60% compared to the mid-knot method and by approximately 14.38% compared to Pagani et al. method[4]. When the specified approximation accuracy is 0.01, the number of feature points required by our method is reduced by approximately 16.81% compared to the mid-knot method and by approximately 15.85% compared to Pagani et al. method[4]. When the specified approximation accuracy is 0.1 and 0.01, the number of feature points required by our method is less than that of the mid-knot method and Pagani et al. method[4], and a comparable approximation accuracy is obtained with the fewest feature points. Since the number of control points obtained by inverse equation (4) is consistent with the number of feature points, the number of control points of the swept surface generated by our algorithm is also minimized under similar approximation accuracy. In addition, the maximum feature point selection multiple set in Table Ⅱ is 20. When the maximum selection multiple is reached, the number of samples in Table Ⅲ where the approximate surface generated by each method still does not reach the specified approximation accuracy is shown. Our method has significantly fewer non-converged samples than the mid-knot method and Pagani et al. method[4], indicating that our method has stronger stability.

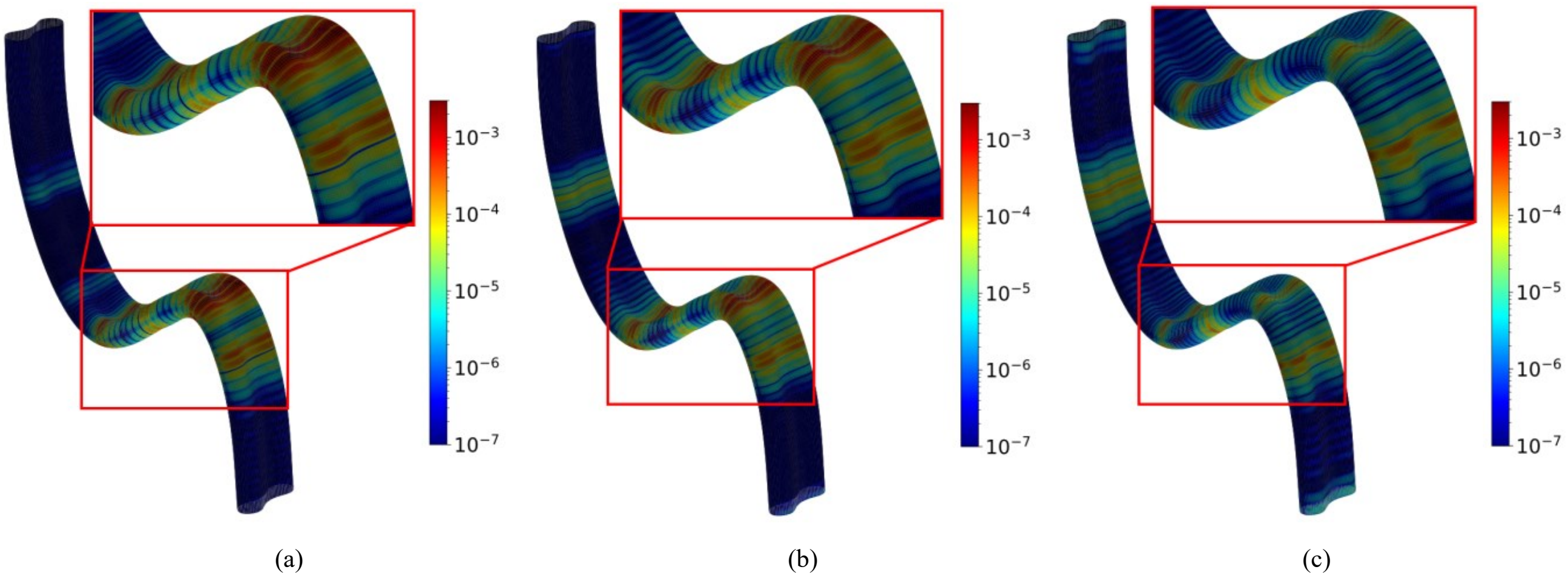

Figure 7. The error of each method on the pipe example (specified selection multiple is 10) :(a) Mid-knot (error: 0.0033); (b) Pagani et al[4] (error: 0.0015); (c) Ours(error: 0.0008).

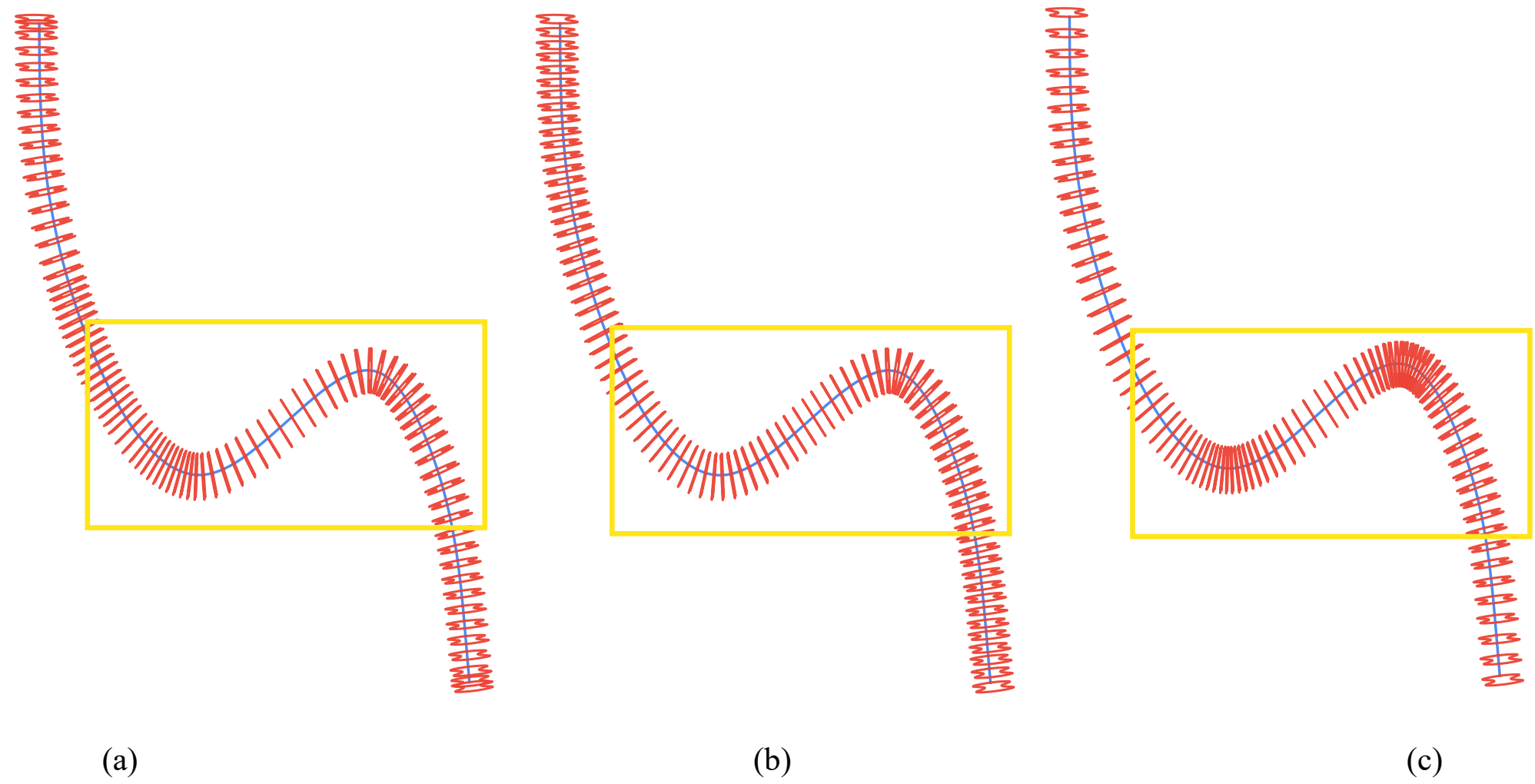

Figure 8. Distribution of cross-sections of each method on pipe example (specified selection multiple is 10) :(a) Mid-knot; (b) Pagani et al[4]; (c)Ours.

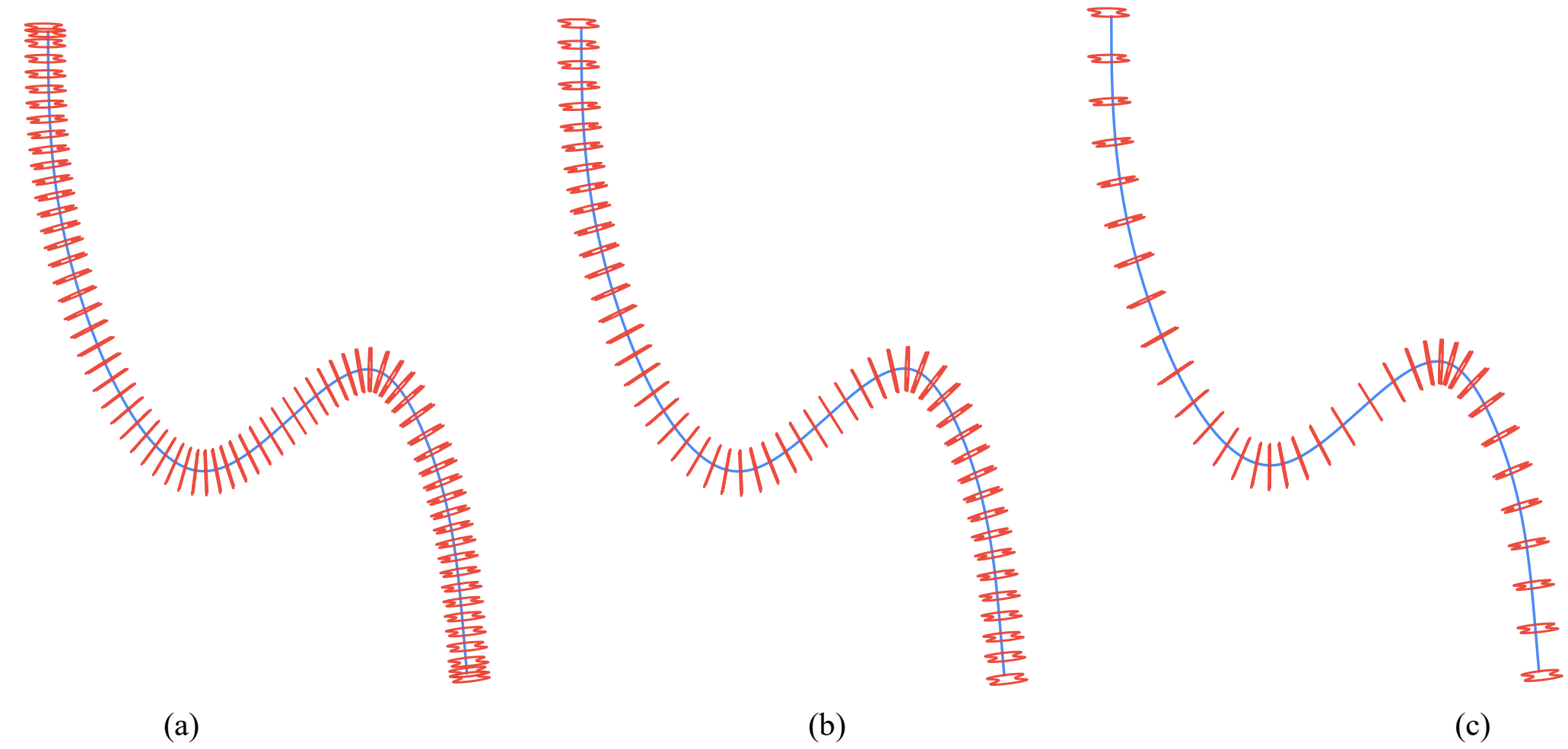

Figure 9. Distribution of cross-sections of each method on pipe example (the surface approximation accuracy is specified as 0.01) : (a) Mid-knot (error: 0.0075, selection multiple: 8); (b) Pagani et al[4] (error: 0.0079, selection multiple: 6); (c) Ours(error: 0.0050, selection multiple: 4).

TABLE III. THE NUMBER OF NON-CONVERGENT SAMPLES OF EACH METHOD UNDER DIFFERENT ACCURACY REQUIREMENTS

| Specified accuracy | Mid-knot[2] | Pagani et al[4] | Our method |
|---|---|---|---|
| 0.1 | 100 | 7 | **1** |
| 0.01 | 193 | 31 | **19** |

Figure 9 shows the position distribution of the three methods for placing the upward cross-section in the pipe example when the specified accuracy is 0.01. Under the same approximate accuracy requirement, the mid-knot method and Pagani et al. method[4] achieve the required accuracy when the selection factor is 8 and 6 times, respectively. Our method achieves a comparable surface approximation accuracy when the selection factor is 4 times, and the number of feature points used is much lower than the other two methods. The results in Figures 6-9 together illustrate the effectiveness of the method in instance construction.

### D. *Effectiveness analysis of weight adjustment*

To verify the effectiveness of the dynamic weight adjustment method, we compared the performance of our algorithm with and without weight adjustment in 969 samples. The results are shown in Table Ⅳ.

TABLE IV. THE AVERAGE ERROR OF WHETHER OUR ALGORITHM PERFORMS WEIGHT ADJUSTMENT

| Selected multiplier | No weight adjustment | Adjust weights |
|---|---|---|
| 1 | 1.903 | **1.644** |
| 3 | 0.917 | **0.774** |
| 5 | 0.404 | **0.321** |
| 7 | 0.158 | **0.133** |
| 10 | 0.039 | **0.036** |

As shown in Table Ⅳ, under different selected multiplier, the effect after weight adjustment is better than the effect without weight adjustment, which demonstrates the effectiveness of the weight adjustment method in our method.

## V. CONLUSION

To achieve a high-accuracy approximation of the swept surface and make more effective use of the known swept surface information, we propose a non-uniform B-spline optimization method for generating the swept surface. First, the geometric features of the swept surface are represented by the surface area, discrete curvature, first derivative, and first derivative rotation angle of the swept surface, and a distribution function is established to measure the irregularity of the surface. At the same time, the weights are adjusted for different samples. Then, feature points are selected based on the distribution function, and the control points of the approximate surface are obtained through inversion calculation. Finally, the feature points are increased by estimating the error of the approximate surface, thereby generating a swept surface that meets the specified approximation accuracy. Experimental results show that our method can effectively improve the approximation effect of the swept surface. Compared with the commonly used mid-knot method and Pagani et al. method[4], our method has better accuracy when the number of specified feature points is large. It can generate a swept surface with considerable approximation accuracy with fewer feature points and has better stability.

## ACKNOWLEDGMENT

This work is supported by the National Natural Science Foundation of China (No. U22A2034, 62177047, 62302530), Key Research and Development Programs of Department of Science and Technology of Hunan Province(No. 2024JK2135), Major Program from Xiangjiang Laboratory (No. 23XJ02005), the Scientific Research Fund of Hunan Provincial Education Department (No. 24A0018), Hunan Provincial Natural Science Foundation (No. 2023JJ40769) and Central South University Research Programme of Advanced Interdisciplinary Studies(No. 2023QYJC020).

## REFERENCES

[1] XU BAI. Research on technologies of offset and sweep based on discrete data. Nanjing: Nanjing University of Aeronautics and Astronautics, 2015(in Chinese)

[2] PIEGL L, TILLER W. The NURBS book[M]. Berlin: Springer, 1997: 472-485.

[3] COQUILLART S. A control-point-based sweeping technique[J]. IEEE Computer Graphics and Applications, 1987, 7(11): 36-45

[4] PAGANI L, SCOTT P J. Curvature based sampling of curves and surfaces[J]. Computer Aided Geometric Design, 2018, 59: 32-48

[5] WANG W, JÜTTLER B, ZHENG D, et al. Computation of rotation minimizing frame in computer graphics[R]. Tech. rep, 2007.

[6] POTTMANN H, WAGNER M G. Contributions to motion based surface design[J]. International Journal of Shape Modeling, 1998, 4(03n04): 183-196

[7] KLOK F. Two moving coordinate frames for sweeping along a 3D trajectory[J]. Computer Aided Geometric Design, 1986, 3(3): 217-229

[8] CHUNG K, WANG W. Discrete moving frames for sweep surface modeling[C]//Proceedings of pacific graphics. 1996, 96: 159-173.

[9] BLOOMENTHAL J. Calculation of reference frames along a space curve[J]. Graphics gems, 1990, 1: 567-571

[10] SILTANEN P, WOODWARD C. Normal orientation methods for 3D offset curves, sweep surfaces and skinning[C]//Computer graphics forum. Edinburgh, UK: Blackwell Science Ltd, 1992, 11(3): 449-457

[11] POSTON T, FANG S, LAWTON W. Computing and approximating sweeping surfaces based on rotation minimizing frames[C]//Proceedings of the 4th International Conference on CAD/CG. 1995

[12] GUGGENHEIMER H. Computing frames along a trajectory[J]. Computer Aided Geometric Design, 1989, 6(1): 77-78

[13] [BLOOMENTHAL M, RIESENFELD R F. Approximation of sweep surfaces by tensor product NURBS[C]//Curves and surfaces in Computer Vision and Graphics II. SPIE, 1992, 1610: 132-144

[14] MICHEL D, ZIDNA A. A new deterministic heuristic knots placement for B-spline approximation[J]. Mathematics and Computers in Simulation, 2021, 186: 91-102

[15] JUNJIE LI. Research on the offset theory of curve/surface and ITS application in the intelligent design of die and mould. Shanghai: Shanghai Jiao Tong University, 2020(in Chinese)